# Agreements between Enterprises Digitized by Smart Contracts in the Domain of Industry 4.0

Kevin Wallis, Jan Stodt, Eugen Jastremskoj and Christoph Reich

Furtwangen University of Applied Science, Germany

## ABSTRACT

*The digital transformation of companies is expected to increase the digital interconnection between different companies to develop optimized, customized, hybrid business models. These cross-company business models require secure, reliable, and traceable logging and monitoring of contractually agreed information sharing between machine tools, operators, and service providers. This paper discusses how the major requirements for building hybrid business models can be tackled by the blockchain for building a chain of trust and smart contracts for digitized contracts. A machine maintenance use case is used to discuss the readiness of smart contracts for the automation of workflows defined in contracts. Furthermore, it is shown that the number of failures is significantly improved by using these contracts and a blockchain.*

## 1. INTRODUCTION

The digital transformation of companies is expected to increase the digital interconnection between different companies to develop optimized, customized, hybrid business models. These cross-company business models require secure, reliable, and traceable logging and monitoring of contractually agreed information sharing between machine tools, operators, and service providers. With blockchain technology, business processes can be accelerated, automated, and secured, opening up new value-added opportunities in the context of digitalization. This is done based on the blockchain key features like immutability, distributed nodes, no need for a trusted third party, self-execution, and accuracy. Other technologies e.g. a central database with application programming interface (API) or a trusted third party often lack some of these capabilities. The central database is managed by a single enterprise, which gives the enterprise a decisive advantage in a case of a dispute. Even with a trusted third party (e.g. a lawyer), an unbiased decision cannot be ensured. Without blockchain, a basis of trust must always be created before awarding a contract to a service provider. This is necessary because the requirements laid down in the contract can usually only be checked to a limited extent or not at all (e.g. due to lack of logging, quality control, monitoring, etc.). The use of a blockchain does not require such a basis of trust, because the quality controls contained in the contract must be stored in the blockchain. Thus, a company can change a service provider without relying on a basis of trust or having to create a new basis of trust [1]. Contract compliance between companies can be enforced by 1) collecting contract relevant data, 2) pushing it into the blockchain, and 3) evaluating it by smart contracts. For example, machine manufacturers who give several years of warranty, would like to have more trust in how their customers are using the machines (e.g. is the machine always running at its limit? Is periodic maintenance adhered to?). Our paper shows different digitizable agreements and uses a maintenance use case to demonstrate the benefits of digitized contracts. Furthermore, challenges and solutions for digitized contracts are shown. Section 2 describes related work based on blockchain and smart contracts. Especially the first approaches of integrating smart contracts in Industry 4.0 use cases. Operational requirements for hybrid business models like a) continuous chain of trust b) digitized contracts and c) data privacy and data governance for digitized contracts are given in Section 3. The differentiation between service level agreements, process level agreements, and business level agreements are done in Section 4. Section 5 contains a table with challenges and solutions for using digitized contracts. A maintenance use case is explained in Section 6 and used in Section 7 to show the improvements which can be achieved by using blockchain and smart contracts in comparison to traditional paper contracts. Section 8 concludes with a summary of our work.

## 2. RELATED WORK

Integrating a blockchain into Industry 4.0 replaces existing error-prone procedures with software centered and documented processes [2][3]. An architectural approach for integrating a blockchain into Industry 4.0 was introduced by [4]. It proposes to use smart contracts to control resources in the production process. IoTChain [5], a blockchain security architecture, combines the adaptive communication environment (ACE) as an authorization framework and the object security architecture for the internet of things (OSCAR) as an encryption framework for the application layer payload. Additional papers that target blockchain in Industry 4.0 are focusing on preserving the privacy of data. Rahulamathavan et al. [6] use decentralized attribute-based encryption and decryption for accessing sensor values. Another access control based on smart contracts is introduced by [7] and uses different contract types like a) access control contracts for specifying access control of multiple subject-object pairs, b) judge contracts for evaluating user's misbehavior during access control and c) register contracts for managing the other contracts. Despite the opportunities for blockchain and Industry 4.0, a profound understanding of the blockchain is essential otherwise a serious financial loss can happen [8].

While blockchain can heavily utilize the advantages of industry 4.0, many enterprises do not possess a high degree of interconnected machines and infrastructure [9].

Still, blockchain provides benefits to companies with a lower degree of utilization of industry 4.0 technology. As shown by Mushtaq et al. [10] their use-cases that provide greater transparency for consumers, a better Product Life Cycle Tracking and Tracing, and a higher degree of automation in terms of communication by the utilization of blockchain.

Following the digitization of information of all kinds, the so-called blockchain technology is currently being used to lay the foundations for the digitalization of trust, monetary values, and services using decentralized architectures. Beside of Bitcoin and other financial services, most blockchain work is presently found in the area of the supply chain. For example, in the food industry, strict environmental control during transportation has to be ensured [11][12] or similar applies to medical products [13].

There are first approaches to use blockchain technologies for smart contracts in Industry 4.0. [14] quite fundamentally describes the ideas that lie behind the present project proposal but is quite strongly oriented towards conventional business processes and not to the connection of machines. A web API for service level agreement (SLA) contracts was introduced by [15]. The work focuses on an API specification for the orchestration of SLA contracts but disregards Industry 4.0 use cases where machines, sensors, etc. are part of the SLA contract. Beside Industry 4.0 [16] discuss the use of smart contract SLAs for mobile communications providers.

# 3. OPERATIONAL REQUIREMENTS FOR HYBRID BUSINESS MODELS

Conceptional, there are three major operational requirements for hybrid business models: first, a continuous chain of trust, second digitizing contracts, and third controlling data exchange between companies (data governance), as depicted in the conceptional framework (see Figure 1). The legal view of smart contracts [17], with the distinction between strong and weak smart contracts and the lexical semantics, is not further discussed in this paper.

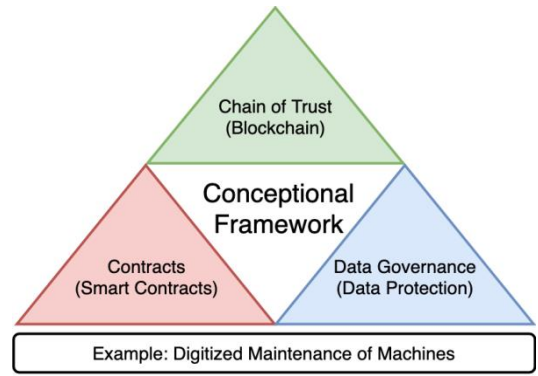

Figure 1. Conceptional Framework

## 3.1. Continuous Chain of Trust

Intrinsically a blockchain can play the role of a chain of trust between the heterogeneous partners. The blockchain technology (e.g. Hyperledger [18], Ethereum) ensures as a decentralized database an encrypted, unchanging, and permanent storage of cross-company information with very high integrity. In companies, such data is usually referred to as audit-proof. Part of the blockchain concept is the technique of distributed consensus building, which replaces trust in a third party with trust in a collective of participants, technology, and cryptography. This enables the realization of novel agile business models that rely on reliable, unchanging information (e.g. contracted metrics). To avoid breaking the chain of trust, the unique identity is essential to prevent the execution of transactions on behalf of another. This identity must be guaranteed for all parties (persons, organizations, machines) and devices (sensors, actors, manufacturing machines, etc.) and must directly be integrated into the blockchain to avoid breaking the chain of trust.

## 3.2. Digitized Contracts

A digitized contract also referred to as a smart contract or chain code, is defined by Clack et al. as "an agreement whose execution is both automatable and enforceable. Automatable by computer, although some parts may require human input and control. Enforceable by either legal enforcement of rights and obligations or tamper-proof execution" [19]. This is a broad definition because they combine the two different smart contract categories: smart contract code and smart legal contracts from Stark [20].

## 3.3. Data Privacy and Governance for Digitized Contracts

When information is exchanged between companies, it requires not only contracts but also security mechanisms that enable the company to retain control of its data at all times, thus protecting their data privacy. Data governance is usually a written document that describes requirements for the proper management of a company's digital data. This policy may include policies for privacy, business process management, security, data quality, etc. For this purpose, additional data governance guidelines (policies) are formulated, that could also be implemented as smart contracts (smart data protection contracts). A smart data protection contract could be formulated, for example, so that only explicitly allowed sensor values from machine A are passed on from company B to company C. This enables real-time control and potential miss-configurations or process errors to be detected promptly.

The implementation of these smart data protection contracts must comply with two constraints to provide the desired protection. First, the aspect confidentiality of privacy [21] cannot be enforced via smart contracts, since data to be examined must be distributed to all or a restricted group of blockchain participants. But since potential confidential data is distributed before confidentiality enforcement, confidentiality is not enforced. A possible solution would be to execute the smart data protection contracts within a secure enclave [22]. Second, the aspect of integrity might be affected by confidentiality enforcement. Providing confidentiality of smart contract execution via a secure enclave may introduce the vulnerability against rollback attacks, harming the integrity [22].

## 4. Hybrid Business Models need Digital Contracts

Typically, a contract is a legal document that defines an agreement between business partners and outlines the services provided, the cost, the resources, etc. Hybrid business models are built by several business services provided by several parties. To get a satisfactory service for the customer, the agreed quality of services between the business partners has to be digitized, to enable automatic monitoring, compliance verification, and initiation of actions, in case contracts are violated.

A business contract is a legally binding agreement between two or more persons or entities. As shown in Figure 2 there are different agreements at various management levels.

**Service Level Agreement:** specifying the quality of service at the IT operation level, which is measured and reported against criteria of technical infrastructures (e.g. bandwidth).

**Process Level Agreement:** specifying the quality of service at the process operation level, that is measured and reported against the context of business processes (e.g. production line processing time).

**Business Level Agreement:** specifying the quality of service at the business operation level, that is measured and reported against the context of business results (e.g. the number of produced workpieces).

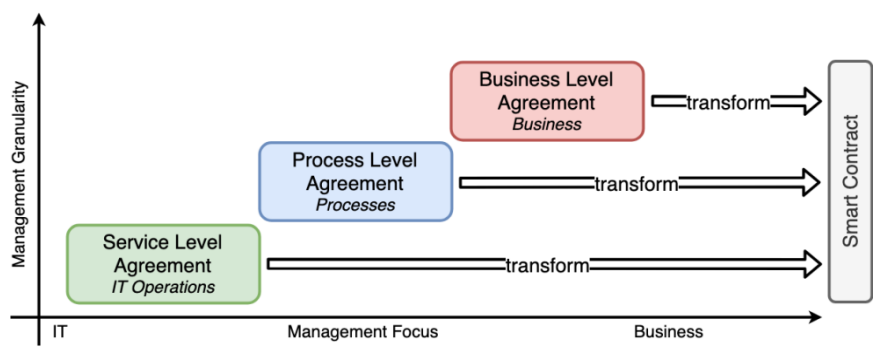

Figure 2. Agreements transformed into Smart Contracts

Sophisticated reporting mechanisms are usually sufficient to document agreements retrospectively. However, to support the progressive digitization of business processes, do real-time reporting, and launching appropriate actions, new approaches are needed to meet the near real-time requirements. Smart Contracts allow a) to model dependencies between the services at the various agreement levels b) to document comprehensibly and unchangeably the specified quality of services of arbitrarily complex systems and c) to monitor specified metrics and activities (workflows) and trigger actions if desired.

# 5. CHALLENGES AND SOLUTIONS FOR SMART CONTRACTS IN INDUSTRY 4.0

Smart contracts play a central role in the digitization of industry 4.0 use cases. In addition to the advantages such as non-repudiation, traceability, and transparency, several challenges are listed in Table 1.

Table 1. Challenges and Possible Solutions for Digitized Contracts in Industry 4.0 Use Cases

| Challenge | Possible Solution |
|---|---|
| Not all parts of a contract between enterprises can be digitized. For example, qualitative measurements, like check cleanliness of a machine, is not possible or too costly to ealize. | Sensors with machine learning, which can measure such qualitative values. |
| Paper-based maintenance contracts may feature a level of condition ambiguity not suitable for a direct transformation [23]. | Evaluate quantifiable values, instruct the human worker to execute the task, and provide proof via a photo, barcode, RFID sensor, or entry of a serial number. |
| Paper-based maintenance contracts may feature certain ambiguous phrases leading to a broad scope of interpretation [17]. | Provide detailed descriptions and definitions of ambiguous phrases. Involve contract partners in smart contract development. |
| Human manual tasks are difficult to integrate. How to verify, that the task has been achieved? | Sensors for checking the result, e.g. spare part replacement is verified by an RFID sensor. |
| The identity of the blockchain participants is costly to be verified. For example, a sensor, that is delivering important information has to be cryptographically identified and integrated into the blockchain. | Usage of a gateway, which is responsible for the communication between sensor and blockchain. The gateway has to provide different features like a cryptographic module, multiple interfaces for sensors, etc. |
| The transfer of data between enterprises is always a source of an unwanted data breach. | A non-disclosure agreement between the different enterprises and encrypted and signed data. |
| Data confidentiality cannot be enforced by a smart contract [22]. | Move confidentiality enforcement from inside a smart contract into an external module or protect entire blockchain peers against confidentiality breaches via secure enclave [22]. |
| The integrity of the data to be recorded on the blockchain must be validated and ensured. | Validation and ensuring data integrity via smart contracts. If a smart contract must be executed within a secure enclave to ensure confidentiality, complete blockchain peer must be executed within a secure enclave [22]. |
| Smart contracts itself can be badly written and therefore is a security thread by themselves [24] [8]. | A validation system, which validates each value before it is stored inside the blockchain. Special caution is required because the validation system can be a single point of failure. |

## 6. MAINTENANCE AS A HYBRID BUSINESS MODEL

Machine maintenance describes the process of keeping up the functionality of machines to ensure flawless and smooth production. Maintenance can be derived in a multitude of variations[1], making a distinction between preventive and corrective maintenance. Preventive maintenance aims to prevent failure of machines by regular performed checks and replacement. On the other hand, corrective maintenance is applied when a machine is broken down and needs to be repaired. These two types can be derived further as described by Nui et al. [25] who outline a more fine-grained sort of maintenance (see Figure 3). Finally, a maintenance type that is often overlooked is the improvement where a machine gets improved by replacing parts with more capable ones or adding parts like sensors.

A fundamental prerequisite for the competitiveness of enterprises is an efficient use of their industrial equipment and machines. To achieve the highest possible availability with the lowest possible costs, machine manufacturers or service providers are offering the maintenance of systems to increase availability. To optimize the maintenance and reduce maintenance costs, information, like operating hours of the machine, the age of the machine, how the machine has been used, workpiece material, etc. are used. The typical monthly or quarterly period of maintenance, for example, can be changed to do it depending on the operating hours of the machine. Besides, the logging of the maintenance process is crucial otherwise serious errors could occur due to an error during maintenance.

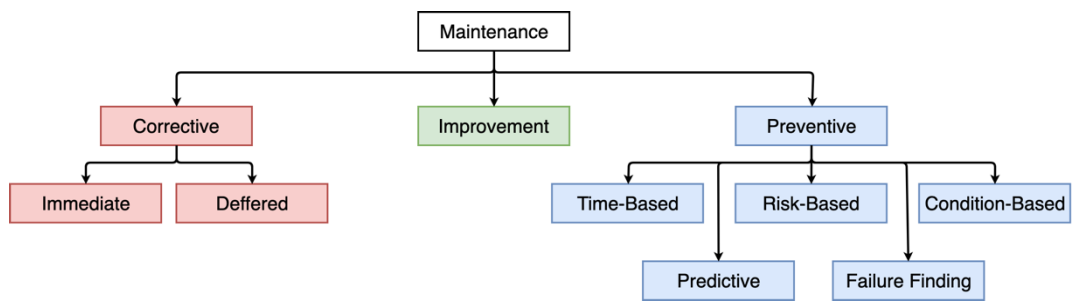

Figure 3. Types of Maintenance

The concept of smart contracts makes it possible to execute predefined processes using rules and execution instructions (small programs) in an automated and decentralized manner.

As seen in Figure 4 there are several stakeholders involved in the maintenance service. There is a machine manufacturer, who build the machine and has the knowledge about the needed maintenance (period time, which component to replace, etc.), the spare part supplier delivers parts to be replaced, the maintenance service provider takes care of maintenance tasks, and the customer (machine user), who has to do standard maintenance (e.g. cleaning once a day). Different user types are handling the machine at the manufacturer's side. The engineers planning the production, the technicians/mechanics repair machines, and the operators do condition monitoring and everyday maintenance, like cleaning.

---

[1] https://www.roadtoreliability.com/types-of-maintenance/

The connections in Figure 4 show the need for possible contracts between the service providers (stakeholders). The goal is to digitize the contracts with smart agreement contracts as good as possible, to monitor in real-time the pre-defined metrics, to show that the interaction between parties is compliant and the protocol is tamper-proof and act immediately if the specified quality of service (metrics) are violated. Typical content of such smart contracts is regular condition analysis of machines, deployment software updates for the machine control, the reaction time in case of failures, repair time of faults, etc.

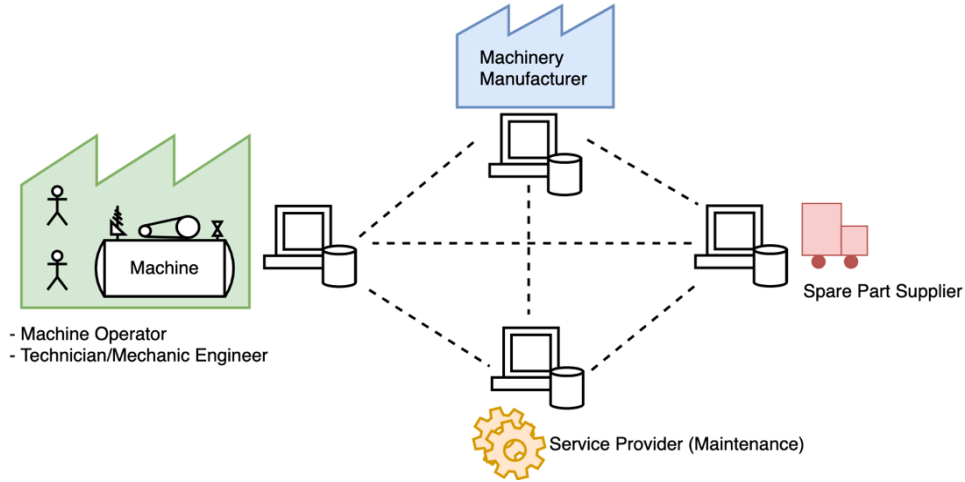

Figure 4. Maintenance Process Stakeholders

## 7. COMPARISON OF PAPER CONTRACTS AND SMART CONTRACTS

In the following, a simplified maintenance use case is used to compare traditional paper contracts with smart contracts. Table 2 describes the steps involved in a traditional maintenance use case and shows which errors can occur. For comparison, Table 3 shows the steps and errors required in an automated maintenance case. In both tables, the first column describes a given task, the second column the interaction between the different task participants, and the last column possible failures for the given task. Possible interaction participants are companies (x, z), employees of the given companies ($e_x$, $e_z$), invoice from company x ($i_x$), checklist from machine m ($c_m$), machine from company x ($m_x$), and blockchain (BC).

Traditional use cases in the simplified case offer 20 possible sources of error and require seven different communication participants. Human failure, in particular, offers a large number of possible errors. Through automation in conjunction with smart contract implementation, the error sources can be significantly reduced [26] (in this example to five).

Table 2. Traditional Paper Contract Maintenance Use Case

| Task | Interaction | Possible Failures |
|------|-------------|-------------------|
| Company $z$ enters a maintenance service contract with company $x$. | $z \rightarrow x$ | |
| An employee $e_z$ performs a check on machine $m_x$. | $e_z \rightarrow m_x$ | - The check was not done<br>- The wrong machine was checked |
| A machine error was found by $e_z$. | $e_z \rightarrow m_x$ | - The error was not found<br>- A wrong error was identified |
| The responsible maintenance service provider $x$ is called by $e_z$. | $e_z \rightarrow x$ | - The maintenance service provider was not called<br>- The wrong maintenance service provider was called<br>- Wrong information about the error was given |
| The maintenance service employee $e_x$ arrives at $z$. | $e_x \rightarrow z$ | - The employee did not arrive<br>- The employee did arrive late |
| $e_x$ inspects $m_x$ via checklist $c_m$. | $e_x \rightarrow m_x$<br>$e_x \rightarrow c_m$ | - The wrong machine was inspected<br>- No error was found<br>- A wrong error was identified<br>- A wrong checklist was used<br>- The checklist was not ticked correctly |
| $e_x$ fixes the error. | $e_x \rightarrow m_x$ | - The error was not fixed<br>- Another error was added |
| $e_x$ documents the error and signs the checklist. | $e_x \rightarrow c_m$ | - The documentation was done wrong<br>- The documentation was not signed |
| Company $x$ sends an invoice $i_x$ to company $z$. | $i_x \rightarrow z$ | - A wrong invoice was sent to $z$ |
| Company $z$ settles the invoice of $x$. | $z \rightarrow i_x$ | - The invoice was not settled |

Table 3. Smart Contract Maintenance Use Case

| Task | Interaction | Possible Failures |
|------|-------------|-------------------|
| Company $z$ enters a smart maintenance service contract with company $x$ and writes it into the blockchain $BC$. | $z, x \rightarrow BC$ | |
| The smart machine $m_x$ detects the error with number 77. | $m_x \rightarrow BC$ | - The machine does not detect the error because the sensor was damaged |
| Company $x$ is informed via $BC$ about the error. | $BC \rightarrow x$ | |
| Company $x$ accepts the maintenance order. | $x \rightarrow BC$ | |
| Maintenance service employee $e_x$ arrives at $z$. | $e_x \rightarrow z$ | - The employee does not arrive<br>- The employee arrives late |
| $e_x$ sets $m_x$ into maintenance mode. | $e_x \rightarrow m_x$<br>$m_x \rightarrow BC$ | |
| $e_x$ fixes the error. | $e_x \rightarrow m_x$ | - The error was not fixed<br>- Another error was added |
| $e_x$ finishes the maintenance. | $e_x \rightarrow m_x$<br>$m_x \rightarrow BC$ | |
| Company $x$ sends an invoice $i_x$ to company $z$ via $BC$. | $i_x \rightarrow BC$<br>$BC \rightarrow z$ | |
| Company $z$ settles the $i_x$ and documents it into $BC$. | $z \rightarrow i_x$<br>$BC \rightarrow x$ | |

# 8. CONCLUSION

The paper has shown a concept of how blockchain and smart contracts can be used to support a well-defined, reliable, traceable interaction of the enterprises to build hybrid business models. It has been shown that the blockchain technology with smart contracts does have great potential to transform the business interaction between companies. Through their intrinsic tamper-proof data storage, their established chain of trust between parties without a central clearing organization, and their possibility to specify contracts for modeling they perfectly fit into the Industry 4.0 domain. Besides, workflow and actions formerly defined in paper contracts between companies can be defined with smart contracts that lead to software supported automatized cross-company interaction processes. Despite the advantages, there are still some challenges (see Section 5) such as lack of total contract digitization, scalability, secure incorporation of external information, data privacy protection, access management, etc.

## ACKNOWLEDGEMENTS

This work has received funding from European Funds for Regional Development (EFRE) and the Ministries for Research of Baden-Württemberg (MWK) in the context of the project BISS 4.0 (biss40.in.hs-furtwangen.de).